\documentclass[preprint2]{aastex}

\shorttitle{Submillimeter Interferometry of Arp~220}
\shortauthors{Wiedner et al.}

\begin{document}

\title{Interferometric Observations of the Nuclear Region of
 Arp~220 at Submillimeter Wavelengths}

\author{M. C. Wiedner}
\affil{Harvard-Smithsonian Center for Astrophysics, 60 Garden St., ms 78, Cambridge, MA 02138, USA}
\email{mwiedner@cfa.harvard.edu}

\author{C. D. Wilson}
\affil{McMaster University, 1280 Main Street West, Hamilton, 
ON L8S 4M1, Canada}
\email{wilson@physics.mcmaster.ca}

\author{A. Harrison\altaffilmark{1} and R. E. Hills}
\affil{Mullard Radio Astronomy Observatory, Cavendish Laboratory,
Madingley Road, Cambridge, CB3 0HE, UK}
\email{harry@biochemistry.ucl.ac.uk, richard@mrao.cam.ac.uk}

\author{O. P. Lay}
\affil{Jet Propulsion Laboratory, California Institute of Technology, 
Pasadena, CA 91109, USA} 
\email{oplay@mail1.jpl.nasa.gov}

\and

\author{J. E. Carlstrom}
\affil{Department of Astronomy and Astrophysics, University of
Chicago, Chicago, IL 60637, USA}
\email{jc@hyde.uchicago.edu}

\altaffiltext{1}{Present address: Biomolecular Structure and Modelling
Unit, Biochemistry and Molecular Biology Dept.,  University College
London, Gower St., London WC1E 6BT, UK}

\begin{abstract}

We report the first submillimeter interferometric observations
of an ultraluminous infrared galaxy. 
We observed Arp~220 in the CO J=3-2 line and 
342~GHz continuum with the single baseline CSO-JCMT 
interferometer consisting of the 
Caltech Submillimeter Observatory (CSO) and the James Clerk Maxwell 
Telescope (JCMT). 
Models were fit to the measured visibilities 
to constrain the structure of the source.
The morphologies of the CO J=3-2 line 
and 342~GHz continuum emission are similar to
those seen in published maps at 230 and 110~GHz. 
We clearly detect a binary source separated by $\sim1\arcsec$ in 
the east-west direction in the 342~GHz continuum.
The CO J=3-2 visibility amplitudes, however, 
indicate a more complicated structure, with evidence for a compact 
binary  at some velocities and rather more extended
structure at others. Less than 30\% of the total CO J=3-2
emission is detected by the interferometer, which implies the presence
of significant quantities of extended gas.
We also obtained single-dish CO J=2-1, CO J=3-2 and HCN J=4-3 spectra. 
The HCN J=4-3 spectrum, unlike the
CO spectra,  is dominated by a single redshifted peak. 
The HCN J=4-3/CO J=3-2, HCN J=4-3/HCN J=1-0 and CO J=3-2/2-1 line ratios 
are larger in the redshifted (eastern) source, which 
suggests  that the two sources may have different physical
conditions.
This result might be explained 
by the presence of an intense starburst that has begun to
deplete or disperse the densest gas in the western source, while
the eastern source harbors undispersed high density gas.

\end{abstract}


\section{Introduction}

\noindent

Ultraluminous infrared galaxies contain extraordinary nuclear starbursts
and, at least in some cases, an active galactic nucleus, all hidden within
a dense shroud of gas and dust. Understanding the mechanisms which
produce and power these luminous galaxies has taken on added urgency with
the discovery of their young counterparts at cosmological distances
\citep{ivison}.
The nearest and prototype ultraluminous infrared galaxy,
Arp~220, is located at a distance of 73 Mpc ($H_o = 75$ km~s$^{-1}$)
and is one of the best studied of this class of galaxies.
The presence of tidal tails observed in the optical 
\citep{arp, joseph}
as well as two compact emission peaks at 
near-infrared \citep{scoville98}, millimeter  
(Scoville, Yun, \& Bryant 1997, hereafter \citet{scoville97};
Downes \& Solomon 1998, hereafter \citet{downes}; Sakamoto et
al. 1999, hereafter \citet{sakamoto}), 
and radio wavelengths \citep{becklin,
norris, sopp} suggests that Arp~220 is a recent merger. 
Given the observed correlation between high infrared luminosity
and disturbed optical morphologies indicative of galaxy interactions
\citep{mirabel}, this merger is likely 
responsible for the extremely high infrared luminosity of Arp~220
[$1.4\times10^{12}$~L$_\sun$, \cite{soifer}]. 
There has been much debate as to whether 
the high infrared luminosity is due to a starburst or 
an active nucleus or a combination of the two, both for Arp~220 in
particular and for ultraluminous infrared galaxies in general
\citep{genzel, scoville97,lutz}.
Recent radio studies of Arp~220 provide support for the 
starburst hypothesis: the 18 cm flux seen in VLBI 
observations is thought to be emitted by luminous radio supernovae 
\citep{smith} and the effects of these supernovae winds are seen in
X-rays \citep{heckman}. Near-infrared \citep{emerson, rieke,  sturm, lutz} 
and earlier radio observations \citep{condon, sopp, baan} 
further support the starburst scenario.

Besides its extremely high infrared luminosity, Arp~220 also 
contains large amounts of dust and molecular gas 
($\sim 9\times 10^{9}M_\sun$ \citet{scoville97}); 
large gas masses of $4-40\times 10^9 M_\sun$ are
typical for ultraluminous infrared galaxies \citep{sanders}. 
The extinction at optical wavelengths is estimated to be at 
least $A_V \sim 50$ and possibly as high as $A_V \sim
1000$ \citep{sturm, downes};
even at near-infrared wavelengths (2.2$\mu$m), 
dust lanes obscure the possible nuclei \citep{scoville98}
These high extinctions mean that 
radio observations are needed to probe the deep interior regions.
\citet{downes} have mapped Arp~220 in CO J=2-1 and 1.3 mm continuum; 
besides detecting two emission peaks,
they also see an extended disk (with 
full-width half-maximum extent of $2\arcsec \times 1.6\arcsec$).
They interpret the emission peaks as the nuclei 
of the merging galaxies embedded in a more extended disk of molecular gas.
\citet{sakamoto} 
refine the model further using their CO J=2-1
and continuum observations; in this model, the nuclei are each embedded in 
their own gas disk which is counter-rotating in the larger common 
disk. An alternative interpretation of the emission peaks
as being due to crowding in the orbits of molecular gas and stars
is given by \cite{eckart}.

This paper presents the very first interferometric submillimeter 
observations of Arp~220 in the CO J=3-2 line
and 0.88~mm continuum.
In fact, it is the first paper to present data from submillimeter 
interferometry of any extragalactic source.
We also present single dish observations of Arp~220
of HCN J=4-3, a high density tracer.
Furthermore, to determine the total CO flux of Arp~220, 
which could be partly 
resolved out by the interferometer, complementary single dish data 
were taken in CO J=3-2 as well as in the CO J=2-1 line.
These submillimeter observations allow us to penetrate deep into 
the interior of the nuclei, while at the same time tracing 
hotter gas, which may be more closely associated with the source of 
the ultraluminous infrared luminosity.
We describe the observations in 
\S~\ref{sec: observations}
and the data analysis in  \S~\ref{sec: modeling}.
The interferometric data were obtained with the single baseline 
CSO-JCMT interferometer, the only currently 
available submillimeter interferometer with sufficient bandwidth
to observe the broad emission lines of Arp~220. With data
from a single baseline, mapping is not possible and so we analyze the
data by making fits in the visibility plane.
The data are compared to published
data at lower frequencies and to single dish data
in \S~\ref{sec: discussion} 
and the conclusions are presented in 
\S~\ref{sec: conclusion}.

\section{Observations and Data Reduction}
\label{sec: observations}

\subsection{Interferometric Data}
\label{sec: int}

\noindent
The interferometric measurements were obtained with the CSO-JCMT 
interferometer on Mauna Kea, Hawaii, consisting of the
Caltech Submillimeter Observatory (CSO) and 
the James Clerk Maxwell Telescope (JCMT)\footnote{The JCMT is operated by
the Joint Astronomy Centre in Hilo, Hawaii on behalf of the parent
organizations Particle Physics and Astronomy Research Council in the
United Kingdom, the National Research Council of Canada and The
Netherlands Organization for Scientific Research.}. 
Usually these two telescopes are operated independently 
for single dish observations, but occasionally they are linked together 
as a submillimeter interferometer with a single baseline of 164~m 
and a minimum fringe spacing of 1.1$^{\prime\prime}$ at
350~GHz \citep{lay94a, lay94b, lay95, lay97}. 
The CSO-JCMT interferometer is currently 
the only submillimeter interferometer with a bandwidth broad enough 
to accommodate the $\sim 800$~km~s$^{-1}$ wide line of Arp~220.

\begin{figure}[htbp]
\plotone{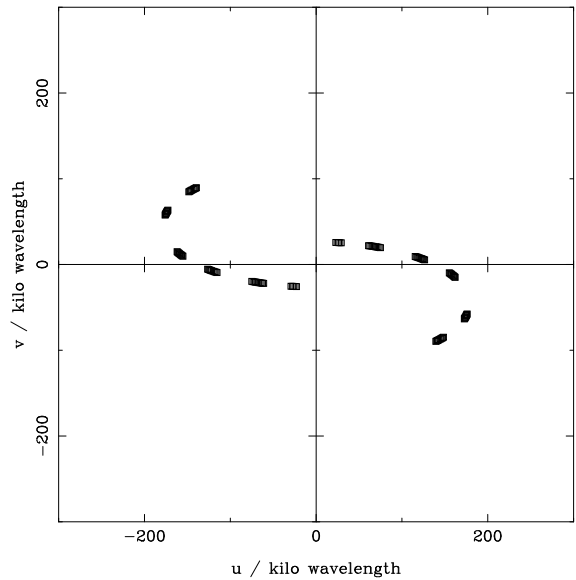}
\caption{Track of CO J=3-2 interferometric observations in (u, v) coordinates.
\label{fig: uv}}
\end{figure}

The $^{12}$CO J=3-2 transition
and associated continuum emission at 342.5~GHz of Arp~220
were observed on 1997 May 10 in good weather with 1.4 mm of 
precipitable water vapor.
The coordinates of Arp~220 used were
$\alpha$(2000)=15$^h$ 34$^m$ 57.$^s$19, 
$\delta$(2000)=23$^{\circ}$ 30$^{'}$ 11$\farcs$3 
(originally $\alpha$(1950)=15$^h$ 32$^m$ 46.$^s$91, 
$\delta$(1950)=23$^{\circ}$ 40$^{'}$ 07$\farcs$9). 
To monitor the gain of the system, 
these observations were interleaved with those of the quasar 3C~345. 
Continuum and line observations of Arp~220 and 3C~345 were alternated
in time. 
We also observed the quasar 3C~273 in both
single dish and interferometric mode to determine the
absolute flux calibration; the interferometric observations were
additionally used to determine the shape of the passband.  
Figure~\ref{fig: uv} shows the (u, v) track of 
the CO J=3-2 data with
a fringe spacing ranging from  $1\farcs2$ to  $5\farcs8$.

Both the Arp~220 and the 3C~345 data were  
divided by a passband created from observations of 3C~273. 
Changes in the relative gain of the interferometer as a function
of time were then calibrated by 
applying the gain curve derived from observations of the point source
3C~345 to the Arp~220 data. Each set of 
ten 10-second integrations was vector-averaged to produce
a series of 100-second integrations with higher signal-to-noise ratios. 

\begin{figure}
\plotone{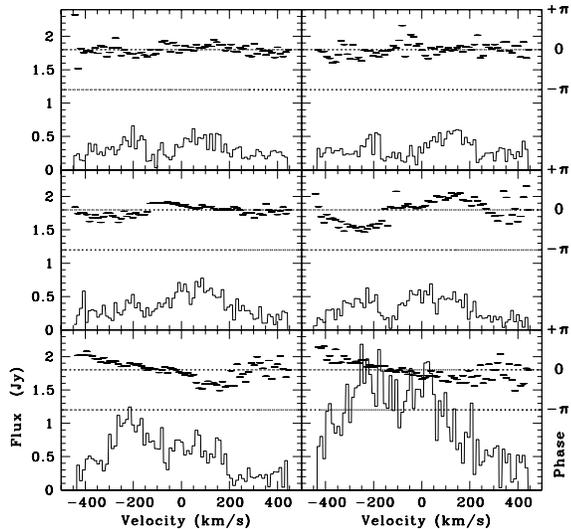}
\caption{Interferometric CO J=3-2  spectra. The histogram shows the
flux on a scale of 0 to 2.4~Jy. The dashes are the phase,
where the upper half of each plot covers the phase from $-\pi$ to
$\pi$. 
The velocities are relative to 
339.5~GHz corresponding to a velocity of 5450~km~s$^{-1}$ in the radio 
definition. 
The first spectrum on the top left was taken at -2.3~h (HA) at (u, v)
coordinates (-140~k$\lambda$,90~k$\lambda$), the top right at HA=-0.8~h
(-170~k$\lambda$,60~k$\lambda$), middle left at HA=2~h
(-160~k$\lambda$,10~k$\lambda$), middle right at HA=3.2~h
(-125~k$\lambda$,-10~k$\lambda$), bottom left at HA=4.5~h
(-70~k$\lambda$,-20~k$\lambda$), bottom right at HA=5.5~h
(-25~k$\lambda$,-25~k$\lambda$).
The continuum
- measured at slightly different (u, v) positions - was extrapolated to the
above (u, v) positions (using our best fitting
binary model) and the expected continuum emissions of 0.16, 0.31 0.28, 
0.15, 0.2 and 0.35~Jy have been subtracted from the respective line data.
\label{fig: spectra}}
\end{figure}

The spectra displayed in Figure~\ref{fig: spectra} were
further averaged to 1000 seconds. 
(The sixth spectrum only contains 500~seconds of data.)
Since the average phase might drift over time scales longer than 100~seconds
due to instrumental or atmospheric phase shifts,  
the average phase over the whole frequency range 
of each 100-second sample was measured and 
then subtracted from the phases in each of the individual velocity channels. 
The resultant line data were then averaged by deriving the arithmetic 
means of the sine and cosine components 
in each individual channel. 

By comparing single dish measurements of Mars with those of 3C~273, we
obtained a flux of $12 \pm 0.5$~Jy for 3C~273, which 
agrees very well with the value of 12.2~Jy measured at 850~$\mu$m with SCUBA 
around the same time \citep{jenness}.
We used the 3C~345 measurements immediately before and after the 3C~273 
data and obtained a 3C~345 flux of 0.90~Jy averaged over the emission
at 342.5 and 339.5~GHz. After averaging and gain calibrating the Arp~220
and 3C~345 data in exactly the same way, the scalar average of the 
3C~345 data was determined and from it the conversion factor from K to
Jy. This conversion may vary between the continuum and CO data and
with velocity bin because the gain
calibration can vary with frequency range.   
We derived conversion factors ranging from
$48$~Jy~K$^{-1}$ to $59$~Jy~K$^{-1}$.
The overall uncertainty in the calibration due to gain 
variations and uncertainties in the planet flux is estimated to be 
20\% \citep{lay94b}. 

Since the continuum data were measured at different (u, v) positions than 
the line data, we extrapolated the continuum data
using the binary model discussed in \S \ref{sec: fitres} and subtracted 
0.16, 0.31, 0.28, 0.15, 0.2 and 0.45~Jy from the six CO
measurements (with increasing hour angle), respectively.
All velocities quoted use the  radio definition 
($v_{rad}=c(\nu_{rest}-\nu_{obs})/\nu_{rest}$) 
with respect to the local standard of rest (lsr).

\subsection{Single-Dish Data}
\label{sec: sd}

\noindent
On 1997 July 17 and 18 we used the 
JCMT to observe emission lines from 
$^{12}$CO J=3-2, $^{12}$CO J=2-1, and HCN J=4-3
under good weather conditions. The observations were  
made with the facility  230~GHz (A2) and  350~GHz (B3) receivers, which
had system temperatures of 270-360~K and 400-600~K, respectively,
in the center of the spectral window.
The spectra were obtained with a chopping secondary mirror using 
a switch cycle of 1~Hz and a beam throw of 60$^{\prime\prime}$ 
in azimuth. We configured the DAS
to give a spectral resolution of 1.25 MHz with a bandwidth of 1.86 GHz for B3
and 0.95 GHz for A2. 
The relative errors in pointing were small, $\sim 2^{\prime\prime}$~rms for
A2 and $\sim 3^{\prime\prime}$~rms for B3, with significant systematic
errors in the pointing apparent when observing at elevations greater
than 80$^o$. Spectra obtained at such high elevations were not used
in the subsequent analysis.

Our data were flux 
calibrated through observations of Mars and Uranus. We found 
significant differences between our derived aperture efficiencies and the 
typical aperture efficiencies for the JCMT. We derive aperture
efficiencies of 0.48 at 230 GHz, 0.42 at 265 GHz, and 0.43 at 345 GHz,
compared to the standard telescope values of 
0.69 at 230 GHz and 0.58 at 345 GHz. We adopt
20\% as the error in the absolute calibration of our single
dish observations. We present all our single dish data on the
T$^*_A$ temperature scale, which is most appropriate for the
compact nuclear emission seen in Arp~220 \citep{scoville97,
sakamoto, downes}.

\begin{figure}[htbp]
\plotone{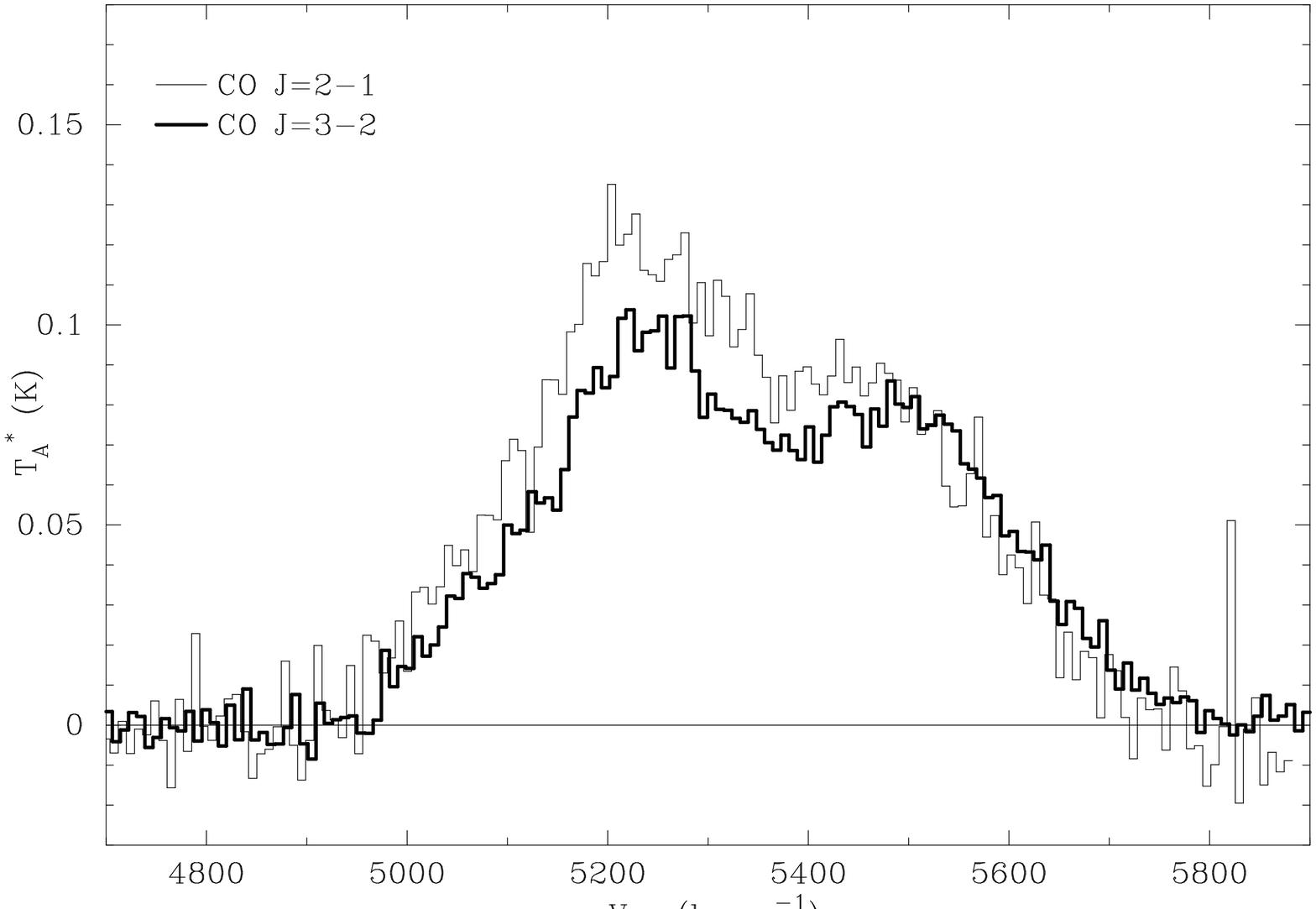}
\caption{CO J=2-1 and CO J=3-2 spectra with a resolution of 22$\arcsec$. 
For CO J=2-1 
the integrated intensity is 53~K~km~s$^{-1}$ or 1730~Jy~km~s$^{-1}$.
For CO J=3-2 
the integrated intensity over the same 22$\arcsec$ beam 
is 45~K~km~s$^{-1}$ or 3700~Jy~km~s$^{-1}$.
\label{fig: compare}}
\end{figure}

We used the package SPECX \citep{padman} to reduce the spectra
and used first-order baselines in all cases.
The CO~J=2-1 spectrum is shown in Figure~\ref{fig: compare}. 
We derive an integrated intensity of 53~K~km~s$^{-1}$, 
which for a conversion factor of 32.6~Jy~K$^{-1}$(T$_A^*$)
corresponds to a flux of 1730~Jy~km~s$^{-1}$. This value is significantly
larger than the flux determined by \cite*{radford} in a
12$^{\prime\prime}$ beam (1040~Jy~km~s$^{-1}$). The interferometric
maps of \citet{scoville97} and \citet{downes} 
show very compact CO J=2-1 emission
with a total extent of roughly $4\times 4^{\prime\prime}$. The larger
flux detected in the 21$^{\prime\prime}$ beam of the JCMT suggests
that there may also be extended CO J=2-1 emission with
roughly half the total intensity seen in the nuclear region. However,
we also obtained CO J=2-1 spectra at positions offset
11$^{\prime\prime}$ (half a beam width) from the central position;
the flux in those offset spectra is roughly half that in the
center, consistent with quite compact emission in the central region.

\begin{figure}[htbp]
\plotone{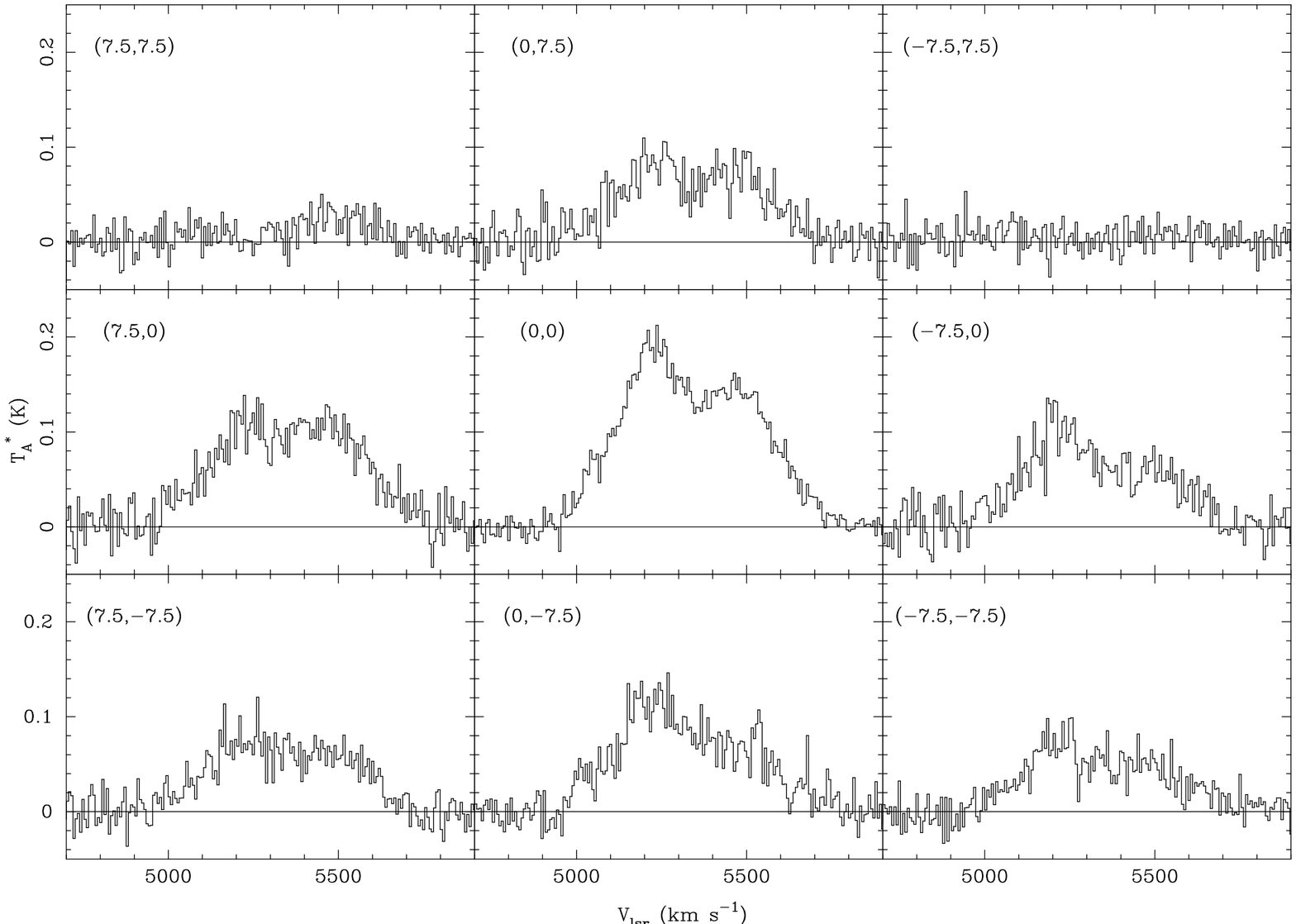}
\caption{Map of CO J=3-2 emission with a resolution of 15$\arcsec$. 
The integrated intensity in the central spectrum is 84~K~km~s$^{-1}$ or
2980~Jy~km~s$^{-1}$
\label{fig: CO32}} 
\end{figure}

A small map of the CO J=3-2 emission is shown in 
Figure~\ref{fig: CO32}.
For the central position we derive an integrated intensity 
of 84~K~km~s$^{-1}$, 
which for a conversion factor of 35.5~Jy~K$^{-1}$ (T$_A^*$)
corresponds to a flux of 2980~Jy~km~s$^{-1}$. When we convolve this map to 
simulate the 21$^{\prime\prime}$ beam of the CO J=2-1 spectrum,
we derive an integrated intensity of 45~K~km~s$^{-1}$ or 3700
Jy~km~s$^{-1}$. 
This integrated intensity is somewhat larger than the value of 
32~K~km~s$^{-1}$ obtained in a similar beam by \cite{gerin}, but
agrees well with the measurement of \citet{mauer99}.
Combining the convolved spectrum with the CO J=2-1  data gives a 
CO J=3-2/J=2-1 ratio of $0.85\pm 0.24$, where the
uncertainty includes the estimated 20\% calibration uncertainty
in each line.

\begin{figure}[htbp]
\plotone{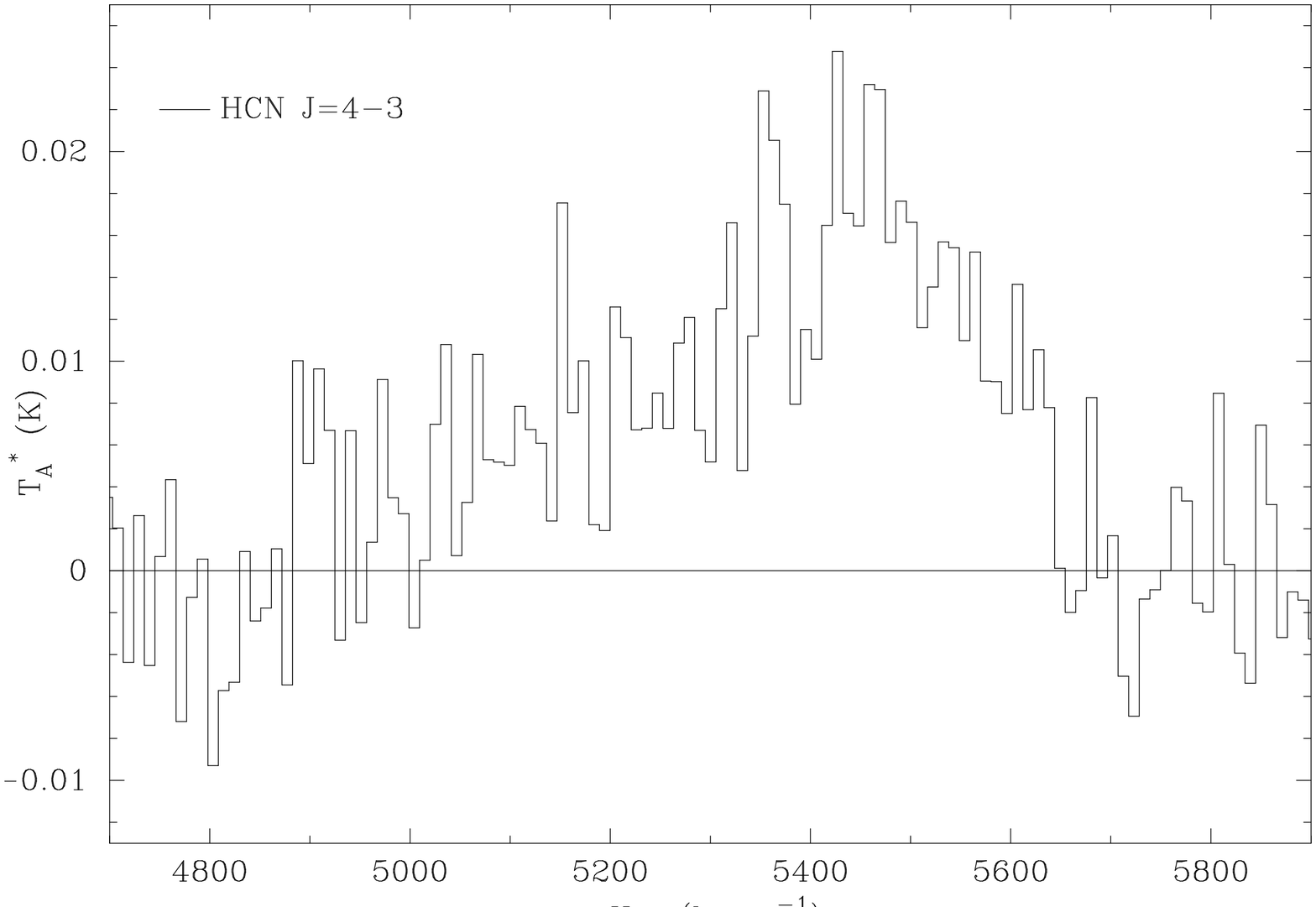}
\caption{HCN J=4-3 with a resolution of 15$\arcsec$. 
The integrated intensity is 7.1~K~km~s$^{-1}$ or 260~Jy~km~s$^{-1}$.
\label{fig: HCN}}
\end{figure}

The HCN spectrum is shown in Figure~\ref{fig: HCN}, 
with an integrated intensity of 
7.1~K~km~s$^{-1}$
(260~Jy~km~s$^{-1}$) for the HCN J=4-3 transition. 
We can combine the HCN and CO data obtained with similar
beams to obtain an HCN J=4-3/CO J=3-2
line ratio of $0.085\pm 0.024$. This line ratio
is quite similar to the value of 0.12
obtained by \cite*{solomon92} 
for the J=1-0 transitions in 
23-28$^{\prime\prime}$ beams.


\section{Analysis of Single Baseline Interferometric Data}
\label{sec: modeling}

The CSO-JCMT interferometer was used to measure 
the cross-correlated amplitude 
and phase of Arp~220 as a function of observing frequency and as 
a function of (u, v) position (Figure~\ref{fig: uv}), 
which is related to the projected baseline.
With a fixed single baseline interferometer, we cannot obtain
sufficient coverage of the (u, v)plane to produce
an image of the source on the sky.
Nevertheless, the (u, v) data themselves contain a lot of information,
which can be extracted by comparing simple models to the data (such as
a single extended disk, two point sources, etc.).
We analyzed the amplitude
and phase data separately following the guidelines in \cite{lay94b}.
The method of fitting models to the visibility {\it{phases}} is explained in 
\S~\ref{sec: theophase}, that of fitting models to the visibility
{\it{amplitudes}} in \S~\ref{sec: theovis} 
and the best model fits to our continuum and CO J=3-2
spectral line data are presented in
\S~\ref{sec: fitres}. Readers who are interested
primarily in the results of our analysis may wish to skip directly to
\S~\ref{sec: fitres}.

\subsection{Fitting Models to Visibility Phase}
\label{sec: theophase}

The interferometer measures the difference in path lengths from the 
source to the  two antennas in terms of phase. When corrected
for geometric and instrumental effects, this phase 
contains information about the source's position in the sky.
Unfortunately, the CSO-JCMT interferometer is not stable enough to 
determine accurately an absolute position of a source from its phase.
However, with our spectral line data, 
it is possible to use the redshifted emission of Arp~220
as a phase reference and to measure the phase of 
the blueshifted emission relative to the redshifted emission.
Fitting models to this phase difference as a function of
(u, v) position can determine the position offset on the sky 
of the blueshifted emission relative to the redshifted emission.
As an example, for a pure east-west baseline, a position offset 
in right ascension (RA) will result in a phase change proportional to
the cosine of the hour angle of Arp~220, whereas an offset in 
declination introduces a phase change proportional to
$\sin({\rm RA})\sin(\delta)$, where $\delta$ is the declination of Arp~220.
More accurately, $\Delta\phi = 2\pi(- u \Delta{\rm RA} - v\Delta \delta)$, 
where $u$ and $v$ are the (u, v) coordinates of the phase reference
source, which 
depends on all three components of the baseline as well as the 
declination and hour angle of the phase reference source \citep*{thompson}.

\subsection{Fitting Models to Visibility Amplitudes}
\label{sec: theovis}

\begin{figure}[htbp]
\plotone{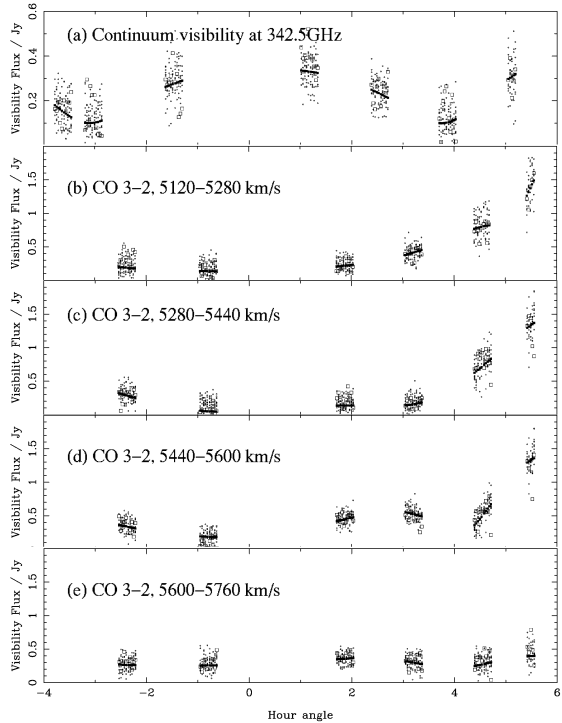}
\caption{Visibilities and model fits to the Arp~220 data. The data are 
represented by squares, the model is the thick line and the dots are 
simulated data given the model and the measured noise. 
(a) Continuum data at 352.5~GHz. (For clarity the
y-axis of the continuum data only runs from 0 to 0.6 Jy in contrast 
to 0 to 2 Jy in all other panels of the line data.) The model represents the
visibility of a binary with two point-like sources with fluxes of 0.25~Jy
and 0.15~Jy, and separated by $1\farcs0$ at a position angle east of
north (P.A.) of
80$^o$.
(b) CO J=3-2 visibilities (after continuum 
subtraction) integrated over 5120 to 5280~km~s$^{-1}$ (radio
definition) and the binary model fit to the data. (Source 1: 3.2~Jy,   
$2\farcs4\times2\farcs4$; Source 2: 0.83~Jy, $1\arcsec\times1\arcsec$;
Separation: $2\farcs8$ at P.A. 30$^o$). 
(c) CO J=3-2 visibilities integrated over 5280
to 5440~km~s$^{-1}$ and binary model fit to the data. (Source 1: 0.65~Jy,   
$0\farcs9\times0\farcs9$; Source 2: 1.0~Jy, $0\farcs95\times0\farcs95$;
Separation: $1\farcs0$ at P.A. 52$^o$).
(d) CO J=3-2 visibilities integrated over 5440
to 5600~km~s$^{-1}$ and binary model fit to the data. (Source 1: 0.55~Jy,   
$0\farcs84\times0\farcs84$; Source 2: 0.92~Jy, $1\farcs0\times1\farcs0$;
Separation: $1\farcs45$ at P.A. 115$^o$).
(e) CO J=3-2 visibilities integrated over 5600
to 5720~km~s$^{-1}$ and binary model fit to the data. (Source 1: 0.3~Jy,   
point-like; Source 2: 0.1~Jy, $0\farcs6\times0\farcs6$;
Separation: $1\farcs7$ at P.A. 130$^o$).
\label{fig: fit}}
\end{figure}

The visibility amplitude changes as a function of time depending on 
the structure of the source and the length and direction of the 
projected baseline. The visibility of a point source, i.e. a quasar, 
will be constant after correction for 
instrumental and atmospheric effects.
Therefore the visibilities of a point source, here 
3C~345, are used to calibrate out any instrumental or atmospheric 
changes. An extended source might be partially resolved out 
around transit, when the projected baseline is the largest, in which
case its visibility amplitude would decrease. The visibility amplitude 
of two point sources will be a minimum
whenever their separation is a multiple of half the fringe 
spacing. From these very simple consideration, it is already 
clear from Figure~\ref{fig: fit} 
that our data are neither consistent with the
visibility expected from a single point source nor are the data in
panels (a), (d) and (e) consistent with that from 
a single extended source. 

Therefore we attempted to fit the Arp~220 data primarily with models
consisting of two sources. 
Each source is assumed to have a Gaussian brightness distribution
of an elliptical shape; this approach results in a model with 10 parameters
(the separation of 
the two sources and the position angle of this separation; 
the flux, minor axis, major axis, and position angle of each source).
It is possible
that models with even more sources would fit the data even better.
However, with measurements of the visibility amplitude at only
6 or 7 different (u, v) positions, we
do not have enough data to constrain models with many more free
parameters.

We used the Maximum Likelihood method to determine which 
source model fit the data best. We first calculated the 
expected visibility amplitude for each model.
Since the noise of the data is known, the probability of each data point 
given the model can be calculated. 
In the case of high signal to noise, the probability is described by   
a Gaussian distribution. However, in our case, the noise is comparable 
to the signal and therefore the Rice distribution describes our data better
\citep{thompson}. The product of the probability of each of the data points 
given the model gives the likelihood of the model. 
The model with the greatest product of probabilities is the most likely.

For the continuum data we then calculated the 
Bayesian error of each of the ten parameters for the most likely model. 
First, the probabilities for each model were normalized by dividing 
them by the maximum probability. 
Then the probabilities of all models with the 
parameter $a$ having a given value $a_1$ were summed. (This sum is
the marginal probability of $a_1$, which is independent 
of the other nine parameters.)
This process was repeated for all the different values 
$a_2$, $a_3$ etc. of $a$.
(If we had a continuous rather than a discrete set of values for $a$, 
this process would be equivalent to integrating the 
probability $p(a,b,c,...,j)$ of the ten dimensional parameter 
space over all parameters but $a$ to give the one dimensional
marginal probability of $a$, $p_{mar}(a) =
\int\!\!\dots\int\!\!\int\! p(a,b,c,\dots,j) \, db \, dc \dots dj$.)
Usually, but not necessarily,  the value $a_{max}$
with the highest marginal probability is also the value found for the 
most likely model.

In a small region around $a_{max}$, the probability distribution 
can be approximated by a Gaussian.
We defined the one sigma Bayesian error of $a_{max}$ as
the distance between $a_{max}$ and $a_{\sigma}$, where 
$a_{\sigma}$ is the location where the 
marginal probability drops to $\exp(-1/2)p_{mar}(a_{max})$. 
(More precisely, $a_{\sigma}$ should define the boundaries 
within which 68\% of the marginal probability lies. 
In the case where the  marginal probability can be approximated by 
a Gaussian, $p_{mar}(a_{\sigma})$ will have the value $\exp(-1/2)
p_{mar}(a_{max})$.
Since in our case the marginal probability distribution of the 
continuum data appears Gaussian, 
we used the latter method to determine the error.)
The error on each parameter gives an indication of how well the 
measured data determine each parameter in a given model, but the 
error contains no information about the validity of the model.

For the line data the Bayesian error could not be calculated, because
the marginal probability is not Gaussian, as there are large secondary 
maxima of the probability. We have therefore taken a different
approach: after finding the most likely model, we {\it{fixed}} all parameters 
but one and defined the 1$\sigma$ error as the distance where the
probability falls to $\exp(-1/2)p(a_{max})$. The difference to
the continuum method is that we do not integrate over all other
parameters, i.e. we are assuming that all other parameters are
accurately known. The advantage of this method is that it ignores
most secondary maxima, as we are only taking a one-dimensional cut in the 
11-dimensional probability surface. Since we are not taking account of 
all other planes, the errors are much smaller and do not reflect the
error due to the existence of secondary maxima. (It might help to 
picture a two-dimensional probability distribution as a landscape with
mountains, where the two dimensional plane represents the two
parameters and the height the probability. The case of the continuum
data can be imagined as one very high mountain in a landscape of
low hills; the line data would correspond to a complicated mountain 
range with many nearly equally high mountains.) 

From the maximum likelihood method, we can only determine 
which of the models we {\it{tried}} is the most likely. 
It is, however, possible that an entirely different model we 
have not considered fits the data much better.
For this reason it is very helpful to have 
maps of the source at other wavelengths (e.g. around 230~GHz), which  
show  the morphology of the source.

\subsection{Results of Model Fitting to Continuum and CO J=3-2 Data}
\label{sec: fitres}

The 342.5 GHz continuum measurements of Arp~220 were fit using both
single and binary source models. A single source model did not fit the
data well.  
Figure~\ref{fig: fit}a shows the visibility data at 342.5~GHz and the
best binary fit to the data. 
The parameters of the best binary fit are listed in 
Table~\ref{tab: cont}, which also gives a comparison to 
published results from continuum data at other wavelengths. Note
in particular 
that, while the continuum data can constrain the strength and separation
of the two sources, the models cannot tell us whether the eastern or the
western source is the stronger one. This limitation is due to the
lack of absolute phase information as discussed in \S~\ref{sec: theophase}.

\begin{deluxetable}{l c c c c c} 
\tabletypesize{\scriptsize}
\tablecaption{Continuum Data}
\label{tab: cont}
\tablewidth{0pt}
\tablehead{
\colhead{} & \colhead{These data}   & \colhead{\citet{sakamoto}} &
\colhead{\citet{downes}} & \colhead{Scoville}   & \colhead{Baan \& Haschick}\\
\colhead{} & \colhead{342.5~GHz}   & \colhead{230~GHz} &
\colhead{230~GHz} & \colhead{2.2$\mu$m}   & \colhead{4.83~GHz}}
\startdata
Morphology & binary & binary & binary & three sources & binary \\
Total Flux (mJy) & 400$\pm$40 & 208 & 175 & \nodata & 214\\
Separation (arcsec)& $1.0 \pm 0.1$ & 0.9 & 0.8 &
        1.13  (NE-W) & 0.98 \\ 
& & & & 1.05  (SE-W) &\\
Position  & $80 \pm 12$ & 100 & 100 & 
        86 (NE-W) & 98\\
~~~~~Angle (deg) & & & & 108 (SE-W) &\\
Eastern Source: & & & & &\\
~~~Flux mJy & 150$\pm$10 & 66 & 30 & \nodata & 88 \\
~~~Size (arcsec$^2$)& $<0.6^2$ & $<0.2^2$ & $0.6 \times 0.6$ & 
$0.34 \times 0.26$ (NE) & $0.42 \times 0.31$ \\
& & & & $0.26 \times 0.23$ (SE) &\\
Western Source: & & & & & \\ 
~~~Flux (mJy) & 250$\pm$40 & 142 & 90 & \nodata & 112  \\
~~~Size (arcsec$^2$) & $<0.6^2 $ & $0.32 \times 0.19$ & 
$0.3 \times 0.3$ & $0.49 \times 0.22$ &
$0.30 \times 0.21$\\
Disk & & & & & \\
~~~Flux (mJy)& 0 & 0 & 55 & \nodata & 14 \\
~~~Size (arcsec$^2$) & \nodata & \nodata & \nodata & \nodata & \nodata \\
Flux cal. error & 20\% & 20\% & 20\% & \nodata & \nodata \\ 
\enddata
\tablecomments{Parameters of the binary model that fits the 
342.5~GHz continuum data best. The errors quoted are the 3$\sigma$
Bayesian error of the model fitting (see \S~\ref{sec: theovis}). The
fluxes have an additional 20\% calibration error. For comparison 
measurements at millimeter \citep{sakamoto,downes},  near-infrared 
\citep{scoville98}, and  radio
wavelengths \citep{baan} are also listed. 
The radio fluxes quoted are from the
naturally weighted 4.83~GHz continuum maps.} 
\end{deluxetable}

\begin{figure}[htbp]
\plotone{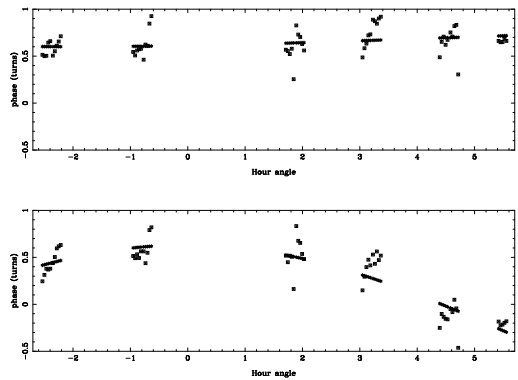}
\caption{
Measurement of source separation using the phase difference between the 
blueshifted (5290-5440~km~s$^{-1}$) and redshifted (5550-5810~km~s$^{-1}$) 
emission of Arp~220. The top graph shows the phase of the redshifted 
emission, the bottom graph shows the phase of the blueshifted  emission.
In both cases, a line fitting the phase of the redshifted emission 
has been subtracted;
the  scatter in the red phase therefore  represents the noise in the
data. The model fit is represented by the thick line and suggests 
an offset of ($0.\arcsec08$,$-0.\arcsec12$) for the redshifted
emission from an arbitrary phase reference and ($-1.\arcsec28$,
$0.\arcsec 18$) for the blueshifted emission.
\label{fig: phase}}
\end{figure}

To determine the relative location of the two CO emission peaks, 
we divided the CO J=3-2 
spectra into two velocity bins. We chose a
velocity range (radio definition) from 5290 - 5440 km~s$^{-1}$ 
for the blue component, and 5550 - 5810 km~s$^{-1}$ for the red
component. We fit a smooth ``gain'' curve to the phase of the red
emission versus time and then subtracted this curve
from both the red and blue phase.
The results of this phase referencing are shown in
Figure~\ref{fig: phase}. We then fit models
to both the red and the blue phase data using the method
described in \S~\ref{sec: theophase}, with the result that
the averaged blue emission originates from
a location offset by $-1.\arcsec36\pm 0.\arcsec14$ in right ascension
and $0.\arcsec30\pm 0.\arcsec14$ in declination
from the averaged red emission.
In the simple case where all of the redshifted emission is emitted
from one point-like component and all the blueshifted 
emission from another, this result gives
the separation of the two components. However, the analysis
of the CO J=3-2 amplitude data described below  shows
that this model of two point sources may be 
an over-simplification.
Therefore, the result from the phase data
may only give a rough estimate of the
separation of the two locations producing most of the CO J=3-2 emission.
From the phase data, it is certain, however, that the redshifted and
blueshifted CO J=3-2 emission originate from different locations
in Arp~220.

Published lower frequency maps show CO J=2-1 and J=1-0 emission with
a complex morphology (\citet{downes}; \citet{sakamoto}),
which may be difficult to model with simple single or binary source models.
To simplify the morphology to some extent, we divided 
the CO data in four velocity bins of 5120 - 5280~km~s$^{-1}$, 5280 - 5440~km~s$^{-1}$, 
5440 - 5600~km~s$^{-1}$ and 5600 - 5760~km~s$^{-1}$, each of which 
corresponds to 
four velocity channels in Figure~20 from \citet{downes}. But even in these narrower
velocities bins, the CO J=2-1 emission is complex. 
In addition, our single (u, v) track cannot sample the more
extended emission adequately, nor does it have much north-south resolution. 

The visibility amplitude plots themselves show some
clear signatures of the source structure.
Figure~\ref{fig: fit}b-d all show an increase in 
amplitude around an hour angle of 5 h, 
which indicates extended emission in 
the east-west direction that was resolved out at longer
baselines. In contrast, the highly redshifted component of the line in 
Figure~\ref{fig: fit}e 
does not show a rise in amplitude as Arp~220 sets,
which means that there is no extended (arcsecond scale)
east-west component at these velocities. The continuum
data in Figure~\ref{fig: fit}a also do not show a 
large increase in amplitude at this hour angle.

\begin{deluxetable}{l c c c c }
\tabletypesize{\scriptsize}
\tablecaption{Best Fitting Models to CO 3-2 Interferometer Data}
\label{tab: CO}  
\tablewidth{0pt}
\tablehead{
\colhead{CO J=3-2} & \colhead{5120-5280}   & \colhead{5280-5440} & 
\colhead{5440-5600} &\colhead{5600-5760} \\
\colhead{Vel. Bin} & \colhead{km~s$^{-1}$}   & \colhead{km~s$^{-1}$} &
\colhead{km~s$^{-1}$} & \colhead{km~s$^{-1}$}}
\startdata
Separation (arcsec) 	& 2.8$\pm$0.3	& 1.0$\pm$0.1	& 1.45$\pm$0.1	& 1.7$\pm$0.3 \\ 	
Position Angle (deg) 	& 30$\pm$6	& 52$\pm$9	& 115$\pm$6	& 130$\pm$12 \\
Eastern Source & & & &\\
~~~Flux (Jy) 		& \nodata & \nodata	& 0.55$\pm$0.15	& 0.3$\pm$0.06 \\
~~~Diameter (arcsec) 	& \nodata	& \nodata &0.84$\pm$0.12 &$<$0.6\\
Western Source & & & & \\ 
~~~Flux (Jy)		& 0.83$\pm$0.18	& 0.65$\pm$0.3	& 0.92$\pm$0.12	& \nodata \\
~~~Diameter  (arcsec) 	& 1.0$\pm$0.3	& 0.9$\pm$0.25 	& 1.0$\pm$0.12 	& \nodata\\ 
Disk & & & & \\ 
~~~Flux (Jy)		& 3.2$\pm$0.6	& 1.0$\pm$0.15 & \nodata	& 0.1$\pm$0.12 \\
~~~Size (arcsec$^2$) 	& 2.4$\pm$1.2 	& 0.95$\pm$0.12 &\nodata	&0.6$\pm$0.6\\
~~~Pos. Angle (deg)	& \nodata & \nodata	& \nodata	& 90 \\
\enddata
\tablecomments{Parameters of models fitting the visibilities in 
four different velocity ranges of CO J=3-2 emission. 
The data do not constrain the models well and
we have chosen the model that is most consistent with the CO J=2-1 maps
of \citet{downes}. The errors quoted here are the 3$\sigma$
errors of each parameter under the assumption that all other
parameters are correct (see \S~\ref{sec: theovis}). There are
additional, mainly systematic, errors discussed in \S~\ref{sec: theovis}
such that the overall error might be as large as a factor of 2.} 
\end{deluxetable}

For the line data (in contrast
to the continuum data), there are many very different models which fit 
each set of visibility amplitudes 
fairly well. Of these models, we have
chosen the one which seemed most consistent with the CO J=2-1 maps
of \citet{downes}.
The visibility data and the model fits are shown in Figure~\ref{fig:
fit} and the parameters of the model are listed in 
Table~\ref{tab: CO}. The errors in Table~\ref{tab: CO} are 3$\sigma$
errors of each parameter assuming all other parameters are accurately
known (see \S~\ref{sec: theovis}). In addition to those errors there
are: (1) an error due to the existence of secondary maxima (i.e. a completely 
different combination of parameters might give a nearly equally 
good fit); (2) an error due to only trying binary models;  
(3) an error due to the uncertainty in attributing the
visibilities to the different features; and (4) the fluxes have an 
additional 25\% calibration error. 

Each model is only one possible way of
describing our data and parameters in Table~\ref{tab: CO} should be 
used with caution. 
The visibility amplitudes are best fit by binaries. 
However, these binaries are not the east-west binary seen
in the continuum data; rather,
it appears to be one or the other of the two continuum nuclei plus
a much more extended CO disk. This larger CO disk appears at
different separations and position angles from the more compact
nuclei in different velocity channels, which complicates
the interpretation of the results.

\begin{figure}[htbp]
\plotone{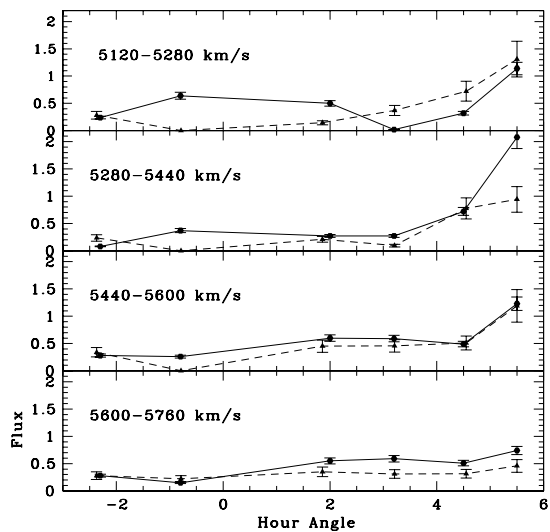}
\caption{
Comparison of CO J=3-2 and CO J=2-1 visibilities. The visibilities were 
calculated for the (u, v) positions shown in Figure~\ref{fig: uv} and for four 
velocity bins of 5120-5280, 5280-5440, 5440-5600 and 5600-5760~km~s$^{-1}$(radio definition) for 
CO J=3-2 (triangles, dashed line) and CO J=2-1 (hexagons, solid line).  
The calibration error on the CO J=3-2 data is estimated to be 25\%, 
that of the CO J=2-1 data 10\%, respectively.
\label{fig: ovro}}
\end{figure}

When comparing the CO J=3-2 emission to maps, it is best to use the
visibilities directly rather than the results from the model fitting.
Thus, we extrapolated the visibilities from the CO J=2-1 OVRO 
data \citep{sakamoto}
to the (u, v) positions observed in CO J=3-2 with the CSO-JCMT 
for a direct comparison, which is shown in Figure~\ref{fig: ovro}.

\section{Discussion}
\label{sec: discussion}

\subsection{Continuum Data}
\label{sec: discont}

The binary model parameters that fit the continuum data best are
listed in Table~\ref{tab: cont}, which also gives  a comparison to continuum
data at other frequencies from the literature. 
The binary separation determined from the continuum data is slightly
larger and the position angle slightly smaller compared
to the two sources seen in the 1.3 mm continuum map of
\citet{sakamoto} and  \citet{downes}.
Our data, however,
agree reasonably well with the position of the formaldehyde emission,
whose peaks are separated by $1\farcs1$ at $79^\circ$
\citep{baan}. Formaldehyde 
is associated with star bursting gas \citep{baan}, 
which will also contain hot dust emitting at submillimeter wavelengths;
it is therefore not surprising that the submillimeter emission shows
the same morphology as the formaldehyde emission.
The two continuum sources are point-like and their size cannot be 
resolved with the CSO-JCMT interferometer. 
From the model fitting (\S \ref{sec: theovis}), the 3$\sigma$
upper limit of $0\farcs6$ 
is consistent with the source sizes 
seen by \citet{downes} and \citet{sakamoto} as well as 
the 2.2 $\mu$m emission, which traces dust-enshrouded young stars
\cite{scoville98}.

We detect only about half of the single dish continuum 
flux [extrapolated with
$\beta=1$ from the 1.1~$\pm$~0.4~Jy total flux, which was  measured by
\cite{eales} at 375~GHz with the JCMT].
This result is slightly surprising because interferometry maps 
do not show significant extended continuum emission at 230~GHz.  
Most of the measured 400~mJy $\pm$ 80~mJy continuum flux is expected 
to come from dust emission rather than synchrotron or free-free emission.
The flux ratio of the two sources is about 1.7,
similar to the 230~GHz data from \citet{sakamoto}, 
which suggests that the continuum flux arises
under similar physical conditions in both sources.
(\citet{downes} divide the flux into 3 sources and so a direct 
comparison is more difficult.)
To first order, the continuum flux is proportional to the product of the 
dust mass and the dust
temperature. Unless the dust temperatures vary by
more than a  factor of 3 between the two nuclei, the mass of dust of the 
two nuclei must be of the same order of magnitude. 

\begin{deluxetable}{lcc}
\tabletypesize{\scriptsize}
\tablecaption{Masses, densities and extinction
\label{tab: calc}}
\tablewidth{0pt}
\tablehead{
\colhead{Frequency} & \colhead{342~GHz}   & \colhead{230~GHz} \\
\colhead{Reference} & \colhead{These Data}   & \colhead{\citet{downes}$^1$}}
\startdata
Eastern Source & & \\
~~~M$_{dust}$ (M$_{\odot}$)   & $2.8 \times 10^7$	& \nodata \\
~~~M$_{gas}$  (M$_{\odot}$)   & $2.8 \times 10^9$  & $0.6 \times 10^9$  \\
~~~N$_{H_2}$  (cm$^{-2}$)   & $>5.5 \times 10^{24}$ & \nodata \\
~~~$\rho_{H_2}$  (cm$^{-3}$)   & $>$15000 & 900-20000\tablenotemark{a}\\
~~~A$_V$	(mag)		  & $>$5800 & \nodata \\
Western Source & & \\
~~~M$_{dust}$ (M$_{\odot}$)   & $4.7 \times 10^7$ & \nodata \\
~~~M$_{gas}$  (M$_{\odot}$)   & $4.7 \times 10^9$ & $1.1 \times 10^9$ \\
~~~N$_{H_2}$  (cm$^{-2}$)   & $>9 \times 10^{24}$ & \nodata \\
~~~$\rho_{H_2}$  (cm$^{-3}$)   & $>$25000 & 900-22000\tablenotemark{a} \\
~~~A$_V$	(mag)		  & $>$9600 & \nodata \\
Disk & & \\
~~~N$_{H_2}$  (cm$^{-2}$)   & \nodata & $10^{24}$ \\
~~~A$_V$	(mag)		  & \nodata & $\sim$1000 \\
\enddata
\tablenotetext{a}{Derived by fitting models of a disk of changing
thickness to CO data.}
\tablerefs{
(1) Downes \& Solomon 1998}
\end{deluxetable}

Gas and dust masses, column densities, volume densities and visual
extinctions are listed in Table~\ref{tab: calc}. To
calculate the dust masses we adopted a temperature of 42~K, which
was derived by \cite{scoville91} by fitting a black body curve to 
IRAS data for Arp~220. We used 
 an opacity coefficient of 1~cm$^{2}$~g$^{-1}$.
The gas mass was calculated using a gas to dust mass ratio of 100
as suggested by \citet{scoville97}. 
Our continuum measurements only give an upper limit of the source 
diameter of $0\farcs6$. We used this value for both the east and 
west source to compute the lower limits of the column densities, 
volume densities  and visual extinctions. 
We used N(H)/E(B-V)=$5.8 \times 10^{21}$ from 
\cite{bohlin} and adopt A$_v$/E(B-V)=3.1
to derive the visual extinction.

The total dust mass derived for our two sources 
of $7.5\times 10^7$ M$_\odot$
agrees fairly well with the estimate of $5\times 10^7$ M$_\odot$ from
the 110 GHz continuum emission by \citet{scoville91}. 
However, our gas mass estimates for the two sources are significantly 
larger than those of \citet{downes} using 230 GHz data; 
moreover they are larger than  
the dynamical masses of $\sim10^9M_\odot$
for these two regions
\citep{sakamoto}. Since the gas mass cannot be larger than the
dynamical mass, one of our assumptions must be wrong.
The assumed gas to dust ratio of 100 is already at the low end of 
ranges typically adopted
and so reducing that ratio seems unlikely. Either a larger grain
emissivity or a larger temperature would act to reduce the dust masses
and hence the gas masses derived from them. 
(Using 100~K versus
42~K will reduce our mass estimates by a factor of 2.6.) 
Changes in grain
properties and elevated temperatures would not be unexpected given
the unusual properties of this starburst region.


%
%
%

\subsection{CO Data}
\label{sec: disco}

The single dish CO J=2-1 and CO J=3-2 data are shown in Figures~\ref{fig:
compare} and \ref{fig: CO32}.
The lines shapes are quite similar for these two transitions and
the measured line ratio of 0.85 agrees quite well with observations
of this transition in normal spiral galaxies and probably
indicates moderately warm (30-50~K) gas \citep{wilson}. 
The CO J=3-2 emission of Arp~220 is clearly moderately
extended, as the flux within a 22$\arcsec$ beam is 20\% larger
than the flux within a 15$\arcsec$ beam.

Our CO J=3-2 integrated intensity agrees quite well with that measured
by \citet{mauer99}
in a similar beam, once both
measurements are converted to the same temperature scale
(i.e. both in T$_{A*}$ or T$_{MB}$). 
There are slight differences in the
line shape between the two observations, particularly in the
relative strength of the red and blue peaks; these
differences are likely due to slightly different pointing between
the two observations, as the rms pointing accuracy in the Mauersberger
et al. data is 5$\arcsec$.

\begin{figure}[htbp]
\plotone{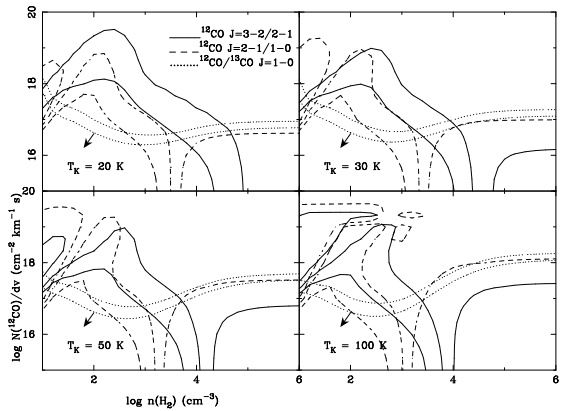}
\caption{Large velocity gradient models for Arp 220 for four different kinetic
temperatures. Three lines are shown for 
the $^{12}$CO J=3-2/2-1 and J=2-1/1-0
line ratios, which correspond to values of $0.84 \pm 0.24$ and 
$0.82 \pm 0.23$, respectively. For the $^{12}$CO/$^{13}$CO J=1-0 line
ratio, two lines are shown, which correspond to the measured 3$\sigma$ upper
limit of 19 from \citet{aalto91} and this value minus 30\% to account for
calibration uncertainty. For this line ratio, an arrow indicates the
direction of the allowed region.  For a given temperature, allowed values of
density and column density occur where the three allowed regions for the
three line ratios intersect. For example, for $T_K = 100$ K, one allowed
region is the triangular region with $n_{H_2} \sim 10^3$~cm$^{-3}$ and
$N(^{12}CO)/dv \sim 10^{17}$~cm$^{-2}$~km$^{-1}$~s, and there is a second
allowed region with $n_{H_2} > 10^4$ cm$^{-3}$ and
$N(^{12}CO)/dv \sim 10^{18}$~cm$^{-2}$~km$^{-1}$~s.
\label{fig: lvg}}
\end{figure}

We can combine our CO data with the CO J=1-0 measurement from
\citet{solomon90}
to obtain beam-matched estimates of the
CO J=3-2/2-1 and J=2-1/1-0 line ratios. These two line ratios
alone do not allow us to place useful constraints on the
physical conditions in the gas using large velocity gradient
models. However, we can get some interesting constraints,
if we include the upper limit to the $^{12}$CO/$^{13}$CO J=1-0
line ratio of 19 from \citet{aalto91}.
One caveat is that
this isotopic line ratio was measured in a much larger beam
(54$\arcsec$) and so may trace emission from beyond the nuclear
region. We compared these three line ratios to the
output from  large velocity gradient models (Figure~\ref{fig:
lvg}) run with a
range of temperatures, densities, and CO column densities
(10-300 K, 10-$10^6$ cm$^{-3}$, $10^{15} - 10^{20}$ cm$^{-2}$
km$^{-1}$ s). Solutions could be found for all the temperatures
we investigated; in general, we can place only a lower limit
on the density for a given temperature, and 
the minimum allowable density decreases as the temperature increases.

The interferometric spectra
are shown in Figure~\ref{fig: spectra} and the visibilities in
Figure~\ref{fig: fit}. 
Figure~\ref{fig: spectra} shows that the shapes of the spectra 
change drastically with (u, v) position. In particular, strong,
mainly blueshifted emission can be seen in the last
spectrum; since the fringe spacing is $5.\arcsec8$, this
emission must be quite extended. In addition,
the visibility phases clearly indicate that there are at least
two sources of emission (Figure~\ref{fig: phase}). 

\begin{deluxetable}{lccc}
\tabletypesize{\scriptsize}
\tablecaption{Comparison of Interferometric CO Observations
\label{tab: compare}}
\tablewidth{0pt}
\tablehead{
\colhead{CO Transition} & \colhead{CO 3-2}   & \colhead{CO 2-1}   &
\colhead{CO 1-0} \\
\colhead{Reference} & \colhead{These Data}   & \colhead{\citet{downes}$^1$}   &
\colhead{\citet{downes}$^1$}
}
\startdata
Separation (arcsec)     & 1.45$\pm$0.1   & 1.3$\pm$0.1   &\nodata \\
Position Angle (deg)& 115$\pm$6    & 85$\pm$6    &\nodata \\
Eastern Source & & & \\
~~~Flux (Jy~km~s$^{-1}$)   & 140$\pm$30   & 220$\pm$44   &\nodata \\
~~~Source diameter (arcsec)  & 0.54$\pm$0.23   & 0.9$\pm$0.1   &\nodata \\
Western Source & & & \\
~~~Flux (Jy~km~s$^{-1}$)   & 380$\pm$60   & 130$\pm$26   &\nodata \\
~~~Source diameter (arcsec)  & 0.97$\pm$0.13   & 0.3$\pm$0.1   &\nodata \\
Disk & & & \\
~~~Flux (Jy~km~s$^{-1}$)   & 690$\pm$90 	& 750$\pm$150   &\nodata \\
~~~Source diameter (arcsec)  & 2.0$\pm$0.9   & 1.8$\pm$0.1   &\nodata \\
Total &&&\\
~~~Flux (Jy~km~s$^{-1}$)   & 1210$\pm$110   & 1100$\pm$220  & 410$\pm$82  \\
\enddata
%
\tablecomments{Averaged parameters of models from
Table~\ref{tab: CO}. (The average source sizes are not the simple 
arithmetic average but are weighted by fluxes.) 
The errors quoted for our data are the 3$\sigma$ errors 
from Table~\ref{tab: CO} propagated according to Gaussian error
analysis. The fluxes have a 25\% calibration error. 
In addition, there are systematic errors discussed in 
\S~\ref{sec: theovis}. The errors of the 
CO J=2-1 and J=1-0 data are taken from D\&S and are expected to represent
the overall error.} 
\tablerefs{
(1) Downes \& Solomon 1998}
\end{deluxetable}

The models fitting the 
visibilities best are listed in Tables~\ref{tab: CO} and~\ref{tab: compare}.
However, the values crucially depend on the model chosen 
and should be used with caution.
The tabulated models 
suggest a total flux of 1210~Jy~km~s$^{-1}$, which is
40\%  of the 3000~Jy~km~s$^{-1}$ 
observed with the JCMT in a $15\arcsec$ beam. 
A lower limit on the flux can be obtained by simply adding
the visibilities in different velocity bins observed for 
the $5\farcs8$ spacing 
which results in 700~Jy~km~s$^{-1}$.
The comparison of single dish and 
interferometer data demonstrates that a significant amount of 
the CO J=3-2 flux is more extended than can be detected with a  
$5\farcs8$ maximum fringe 
spacing, and care should be taken to compare only data with the 
same beam size/fringe spacing.

%
Using the empirical relationship between CO emission and molecular
mass \citep{petitpas}, the average CO J=3-2/1-0 line ratio 
of 0.7 and the fluxes 
determined from the model fitting (Table \ref{tab: compare}) we 
obtained molecular 
masses of $5.3 \times 10^9$ M$_\odot$ for the western source, 
$1.9 \times 10^9$ M$_\odot$ for the eastern source and $9.6 \times 10^9$ 
M$_\odot$ for the disk. These masses agree well with the masses
derived from the continuum emission in \S~\ref{sec: discont}, but are
still larger than the dynamical masses; this result suggests the
CO-to-H$_2$ conversion factor is lower than the value in the Milky
Way, consistent with other results for ultraluminous 
infrared  galaxies \citep{solomon97}.

The CO J=3-2 interferometric data suggest that the western (blueshifted)
source is brighter than the eastern (redshifted) source; the 
single dish spectrum also indicates slightly 
stronger emission in the blueshifted
part of the line. This interferometric result is
in the opposite sense to that seen by \citet{downes}, and could indicated
temperature differences between the two nuclei. [However,
\citet{sakamoto} clearly show the western source is brighter
than the eastern source at CO J=2-1, so perhaps there is a typographical
error in \citet{downes}.]

Morphologically, the CO J=3-2 interferometric data indicate the presence
of two fairly compact emission regions with a more extended disk. However,
the morphology 
appears to be complex. 
In general, our data are consistent with \citet{downes} as well
as \citet{sakamoto} data and small inconsistencies are most likely due to
an over-simplification in our model, which fits only two Gaussian peaks
to the emission. 

Figure~\ref{fig: ovro} shows visibilities of CO J=3-2 and J=2-1 taken
at exactly the same (u, v) positions. The shape 
of the visibility curves roughly agree in the two most redshifted bins,
with larger differences between the two transitions seen in the
two blueshifted bins. In particular,
in the 5280-5440~km~s$^{-1}$ frequency bin, we  
see more extended emission in CO J=2-1 than CO J=3-2. The CO J=2-1 might
also show more evidence for a binary structure. 
The visibilities at HA=-0.8 are the
best representatives of the eastern and western source we have, as 
the disk will be mostly resolved out at this long
projected baseline (-170k$\lambda$,60k$\lambda$, fringe spacing $1\farcs1$).
For this (u, v) point  the
CO J=3-2/J=2-1 line ratios (converted to~K scale) 
are approximately 0.14, 0.12, 0.08 and 0.92 for increasing velocities. 
In the high resolution CO J=2-1 map of \citet{sakamoto},
the western source 
dominates between 5050 and 5450~km~s$^{-1}$, which  
corresponds to our first two velocity bins, 
and the eastern source dominates 
between 5500 and 5650~km~s$^{-1}$, which  corresponds to our
last two velocity bins.
The average line ratio for the blueshifted 
western source is 0.1, significantly below the 
average of 0.5 for the eastern source. 
This results could be interpreted as the eastern source being warmer 
(or denser) than the western source and is consistent with results
from single dish HCN and CO observations discussed in the next section.


\subsection{HCN Data}
\label{sec: dishcn}

Figure~\ref{fig: HCN} shows the HCN J=4-3 spectrum obtained with the
JCMT. 
In contrast to the CO J=3-2 spectrum (Figure~\ref{fig: compare}) 
and the HCN J=1-0 spectrum \citep{solomon92}, 
the HCN J=4-3 spectrum is dominated by a single redshifted emission
peak. 
We divided the HCN emission into red and blueshifted emission 
at 5430~km~s$^{-1}$, where the CO as well as HCN 1-0 spectra have a 
dip in the intensity. 
We converted the HCN J=1-0 line to the T$_A^*$ temperature scale 
using a main beam efficiency of 0.6 \citep{radford}
and scaled the integrated intensity by a factor of four to 
correct for the difference in beam sizes.
Note that the difference in beam size leads to an uncertainty in the 
line ratio because we do not know the true source structure; 
the factor of four scaling is only strictly
appropriate for a point source. 
Our HCN J=4-3 and CO J=3-2 data were obtained with almost identical
beam sizes and so we can calculate a line ratio without the need
to apply any additional corrections. 
The integrated intensities and the line ratios of HCN J=4-3/J=1-0 
and HCN~J=4-3/CO~J=3-2 are listed in Table~\ref{tab: moreHCN}.

\begin{deluxetable}{lccccc}
\tabletypesize{\scriptsize}
\tablecaption{HCN intensities and line ratios
\label{tab: moreHCN}}
\tablewidth{0pt}
\tablehead{
\colhead{Source} & \colhead{HCN J=4-3}   & \colhead{HCN
J=1-0\tablenotemark{a}}   & \colhead{CO J=3-2} & \colhead{HCN J=4-3/J=1-0} &  \colhead{HCN J=4-3/CO J=3-2}  \\
\colhead{} &\colhead{(K~km~s$^{-1}$)} & \colhead{(K~km~s$^{-1}$)}   
&\colhead{(K~km~s$^{-1}$)} & \colhead{}& \colhead{} 
}
\startdata
Eastern Source & 4.0 & 9.6 & 36 & 0.42 & 0.11 \\
Western Source & 3.0 & 10.8 & 48 & 0.28 &  0.064 \\
Total	       & 7.1 & 20.4 & 84 & 0.35  & 0.085   
\enddata
\tablenotetext{a}{The integrated T$_{MB}$ from \cite{solomon92} were
converted into T$_{A*}$ by multiplication with the telescope
efficiency of 0.6 
and corrected to a 14$\arcsec$ beam by multiplying by
a factor of four.}
\end{deluxetable}


The line ratios both in HCN J=4-3/J=1-0 and  
HCN J=4-3/CO J=3-2 are larger for the redshifted eastern source than 
the blueshifted western source. 
Since the HCN J=4-3/J=1-0 line ratio is not affected by abundance 
changes, it seems likely that the physical conditions are different between
the two emission peaks and that the eastern source is either at 
a higher temperature and/or has a higher density.
These conclusions are supported by the work of \citep{aalto02},
who detected CN J=2-1 emission from the blueshifted part of the line,
and HC$_3$N emission in the redshifted part of the line. They
suggest that CN emission is an indicator of a photon dominated
region while the HC$_3$N emission is an indicator of hot cores,
and suggest that the two nuclei may be in different evolutionary
states.
Infrared maps show a dust lane across the eastern nucleus
and a more symmetric morphology in the western nucleus
\citep{scoville98}, which could be at least superficially consistent
with our observed line ratio variations.
\citet{downes} suggest that the western nucleus is undergoing an
intense starburst; if this starburst has
dispersed much of the gas in the western nucleus, while the
eastern nucleus continues to contain a larger mass of gas at high densities,
this scenario would also be consistent with our data.

\section{Conclusions}
\label{sec: conclusion}

We have presented the first interferometric observations of 
Arp~220 at submillimeter wavelengths. 
The interferometric visibilities of the CO J=3-2 line and 342~GHz continuum 
are largely consistent with the emission morphology seen 
previously at lower 
frequencies. We clearly detect continuum and CO J=3-2 emission from at
least two sources separated by $\sim 1\arcsec$ at P.A. $\sim80^o$.
The CO J=3-2 visibility amplitudes show additional extended structure with a
complex morphology. 
Masses, column densities, volume densities and optical extinction 
calculated for both emission sources agree with previous estimates 
within the errors and underline that the center of Arp~220 contains
large amounts of molecular gas ($\sim8~10^9$ M$_\odot$).

Single-dish data indicate that the 
CO J=3-2 emission is moderately extended compared to the 15$\arcsec$
beam of the JCMT. Though the 
continuum visibilities show no sign of an extended source, the single
dish continuum flux is about twice that detected with the interferometer.
In single dish data, the HCN J=4-3/J=1-0, HCN J=4-3/CO J=3-2 and 
CO J=3-2/J=2-1 
ratios are all larger for the redshifted portion of the line 
than for the blueshifted portion.
These observations suggest that the redshifted eastern source 
is denser and/or hotter than the blueshifted western source.
This results could provide support for \citet{downes} hypothesis that the 
western source is currently undergoing an intense starburst that 
has dispersed
the dense gas, whereas the eastern source still
 harbors dense molecular material.

\acknowledgments

We are extremely grateful to Kazushi Sakamoto who extracted a CO J=2-1 
visibility from his data that exactly matched our (u, v) track. 
Furthermore 
we would like to thank David Brown, Claire Chandler and Brenda Matthews 
for help during the CSO-JCMT observations.
The CSO is supported by NSF grant AST 99-80846. The JCMT is operated
by the Royal Observatories on behalf of the Particle Physics and
Astronomy Research Council of the United Kingdom, the Netherlands
Organization of Scientific Research, and the National Research Council
of Canada. MCW was supported by
an HSP III scholar-ship of the German Academic Exchange Service as
well as the Sir Isaac Newton Scholarship.

\end{document}